\documentclass[seceq,preprint]{ptptex}

\usepackage{wrapft}

\usepackage{amsmath,amssymb,amscd,amsbsy,amsgen,amsopn,amstext,amsxtra}
\usepackage[mathscr]{eucal}

\newcommand{\iDsla}{iD\!\!\!\!/\,}

\pubinfo{Vol.~115, No.~6,  June 2006}
\preprintnumber[3cm]{NUP-A-2005-4} 

\title{
Charge Quantization Conditions Based on  \\
the Atiyah-Singer Index Theorem 
}

\author{
Shinichi \textsc{Deguchi}$^{1,}$\footnote{E-mail:  deguchi@phys.cst.nihon-u.ac.jp}  
and Kaoru \textsc{Kitsukawa}$^{2,}$\footnote{E-mail:  kaworu@phys.cst.nihon-u.ac.jp} 
}

\inst{
$^1$Institute of Quantum Science, College of Science and Technology, \\ 
Nihon University, Tokyo 101-8308, Japan
\\
$^2$Graduate School of Quantum Science and Technology, \\ 
Nihon University, Tokyo 101-8308, Japan
}

\recdate{
February 27, 2006}

\abst{
Dirac's quantization condition, $eg=n/2$ ($n \in \Bbb Z$),  
and Schwinger's quantization condition, $eg=n$ ($n \in \Bbb Z$), 
for an electric charge $e$ and a magnetic charge $g$ 
are derived by utilizing the Atiyah-Singer index theorem in two dimensions. 
The massless Dirac equation on a sphere with a magnetic-monopole background is solved 
in order to count the number of zero-modes of the Dirac operator.  
}

\begin{document}
\maketitle

\section{Introduction}

The Dirac quantization condition for an electric charge $e$ and a magnetic 
charge $g$ reads, in units such that $\hbar=c=1$, 
\begin{align}
eg=\frac{n}{2}\,, \quad n \in \Bbb Z \,.
\label{1}
\end{align}
Dirac discovered this condition by noting that 
the phase of the wave function of an electrically charged particle can change 
by only an integral multiple of $2\pi$ under a single rotation of the wave function around 
a so-called Dirac string \cite{Dir, Fel, Shn}. 
Wu and Yang derived the condition (\ref{1}) without reference to the concept  
of a Dirac string \cite{WY, Fel, Shn}. 
They considered two slightly overlapping hemispheres 
(often referred to as ^^ patches' or ^^ charts') that together surround a magnetic monopole, 
and introduced a wave function (or section) for an electrically charged 
particle into each hemisphere. 
Then the condition (\ref{1}) is obtained by noting that 
the phases of the two wave functions, which are connected via a gauge transformation, 
can differ by only an integral multiple of $2\pi$ at a given point in the overlapping region. 
Without explicit use of gauge potentials, Jackiw derived Eq. (\ref{1})  
by examining the associative law for translation operators in the case that a magnetic 
monopole exists \cite{Jac}. 
In this argument, Eq. (\ref{1}) is obtained from the fact that 
the extra phase (or three-cocycle) arising in a nonassociative algebra must be equal to 
an integral multiple of $2\pi$ to maintain the associativity of translation operators.

In this paper we study an alternative approach to deriving Eq. (\ref{1}) using 
the Atiyah-Singer index theorem in two dimensions \cite{AS,ABP,NS,EGH,Nak,FS,Ber}. 
As we see below, in order to obtain Eq. (\ref{1}) in this approach, it is only necessary to 
solve a simple Dirac equation in two dimensions and to formally count  
the number of zero-modes of the Dirac operator.  
In addition to the Dirac quantization condition (\ref{1}), 
our approach yields the Schwinger quantization condition, \cite{Sch, Fel, Shn}
\begin{align}
eg=n\,, \quad n \in \Bbb Z \,.
\label{1-2}
\end{align}
This condition was discovered by Schwinger in his study of 
a relativistic quantum field theory of electric and magnetic charges \cite{Sch}. 
There, it is verified that the relativistic invariance of the theory is maintained 
only when the condition (\ref{1-2}) is satisfied.  
The present paper treats the conditions (\ref{1}) and (\ref{1-2}) in a unified manner;   
each is obtained by fixing a parameter contained in the gauge potential 
at a suitable value.

This paper is organized as follows. 
Section 2 presents the Atiyah-Singer index theorem in two dimensions. 
In \S3, a massless Dirac equation with a magnetic-monopole background is 
solved to find its regular solutions, namely the zero-modes of the Dirac operator. 
In \S4, the number of  zero-modes is counted to derive the charge 
quantization conditions (\ref{1}) and (\ref{1-2}). 
Section 5 contains concluding remarks. 
The Appendix is devoted to a simple proof of the Atiyah-Singer index theorem 
in two dimensions so that this paper may be self-contained.

\section{Atiyah-Singer index theorem in two dimensions} 

Let $\mathcal{M}$ be a compact, oriented, two-dimensional Riemannian manifold 
without boundary. Assuming that $\mathcal{M}$ possesses a spin structure, 
we consider a self-adjoint Dirac operator $\iDsla$ containing  
a hermitian Yang-Mills connection $A$ 
that takes values in the Lie algebra of a compact, simple 
Lie group $\mathcal{G}$. 
The self-adjointness of $\iDsla$ is defined with respect to the natural 
$L^2$ inner product on spinors, which is given by integration over $\mathcal{M}$ involving 
the Riemannian surface-element on $\mathcal{M}$.  
In terms of local coordinates 
$(q^\alpha)$ $(\alpha=1,2)$ on $\mathcal{M}$, the Dirac operator can be 
expressed as  
\begin{align} 
\iDsla \equiv i\sigma_a e_{a}{}^{\alpha} D_{\alpha} \,, 
\label{2}
\end{align}
with 
\begin{align}
D_{\alpha} \equiv  \sigma_0 \partial_\alpha +\frac{i}{2} \omega_{\alpha} \sigma_{3} 
-ieA_{\alpha}\sigma_0 \,. 
\label{3}
\end{align}
Here $\partial_\alpha \equiv \partial/\partial q^\alpha$, 
$e_{a}{}^{\alpha}$~$(a=1,2)$ is an inverse zweibein related to  
the Riemannian metric $g_{\alpha\beta}$ on $\mathcal{M}$, 
$\omega_{\alpha} $ is a spin connection in two dimensions, 
$e$ is an electric charge, and 
$\sigma_0$ denotes the $2\times 2$ unit matrix, while  
$\sigma_a$ and $\sigma_3$ denote the Pauli matrices 
\begin{align}
\sigma_1=\left(\begin{array}{cc}
            0 & 1 \\
            1 & 0 \\
          \end{array} \right) , \quad 
\sigma_2=\left(\begin{array}{cc}
            0 & -i \\
            i & 0 \\
          \end{array} \right) , \quad 
\sigma_3=\left(\begin{array}{cc}
            1 & 0 \\
            0 & -1 \\
          \end{array} \right) .
\label{4}         
\end{align}

Now consider the positive chirality zero-modes  
$\{ \varphi_{0,\nu_{+}}^{+} \}$ $(\nu_{+}=1,\ldots,\frak{n}_{+})$ 
and the negative chirality zero-modes 
$\{ \varphi_{0,\nu_{-}}^{-} \}$ $(\nu_{-}=1,\ldots,\frak{n}_{-})$  
of $\iDsla$, characterized together by 
\begin{align}
& \iDsla \varphi_{0,\nu_{\pm}}^{\pm}(q) =0\,, 
\label{5} \\
& \sigma_3 \varphi_{0,\nu_{\pm}}^{\pm}=\pm \varphi_{0,\nu_{\pm}}^{\pm} \,. 
\label{6}
\end{align}
Here, $\frak{n}_{+}$ ($\frak{n}_{-}$) denotes the number of positive (negative) 
chirality zero-modes. 
Then the Atiyah-Singer index theorem in two dimensions reads 
\cite{AS,ABP,NS,EGH,Nak,FS,Ber}
\begin{align}
\frak{n}_{+}-\frak{n}_{-}=\frac{e}{4\pi} \int_\mathcal{M} d^2 q \,
{\rm tr}\, \varepsilon^{\alpha\beta} F_{\alpha\beta} \,, 
\label{7}
\end{align}
where $\varepsilon^{\alpha\beta}$ $(\varepsilon^{12}=1)$ is the contravariant 
Levi-Civita tensor {\em density} in two dimensions, and $F_{\alpha\beta}$ is the 
field strength of $A_{\alpha}$,  
\begin{align}
F_{\alpha\beta}&\equiv \partial_{\alpha}A_{\beta}-\partial_{\beta}A_{\alpha}
-ie[A_{\alpha}, A_{\beta}] \,. 
\label{8}
\end{align}
The notation ^^ ^^ tr" indicates the trace over the generators of $\mathcal{G}$. 
A simple proof of the Atiyah-Singer index theorem 
in two dimensions is given in the Appendix.

\section{Massless Dirac equation with a monopole background}

Let us consider the case in which $\mathcal{M}=S^{2}$ and $\mathcal{G}=U(1)$. 
Then the Atiyah-Singer index theorem (\ref{7})  reads
\begin{align}
\frak{n}_{+}-\frak{n}_{-}=\frac{e}{4\pi} \int_{S^{2}} d^2 q \,
\varepsilon^{\alpha\beta} F_{\alpha\beta} \,,
\label{7-1}
\end{align}
with $F_{\alpha\beta}=\partial_{\alpha}A_{\beta}-\partial_{\beta}A_{\alpha}$. 
In terms of spherical coordinates, $(q^1, q^2)=(\theta, \phi)$  
$(0\leq \theta \leq \pi,\, 0\leq \phi < 2\pi)$, on $S^{2}$ of radius $r$, 
the diagonal zweibein $e_{\alpha}{}^{a}$ and its inverse $e_{a}{}^{\alpha}$ 
are given by 
\begin{align}
(e_{\alpha a})=\mathrm{diag}(r,\,r \sin\theta) \,,
\quad 
(e_{a}{}^{\alpha})=\mathrm{diag}(r^{-1},\,r^{-1} \sin^{-1}\theta) \,, 
\label{9}
\end{align}
and thus the Riemannian metric takes the standard form 
$(g_{\alpha\beta})=(e_{\alpha a} e_{\beta a})
=\mathrm{diag}(r^2,\,r^2 \sin^2\theta)$. 
The associated spin connection $\omega_{\alpha} $ is found by using  
the torsion-free condition\footnote{The torsion-free condition is given by 
$\varGamma_{\beta\gamma}{}^{\alpha}=\varGamma_{\gamma\beta}{}^{\alpha}$,  
with $\varGamma_{\beta\gamma}{}^{\alpha}$ being the affine connection. 
The first line of Eq. (\ref{10}) is obtained by applying the 
torsion-free condition to Eq. (\ref{A10}) in the Appendix.} and  
Eq. (\ref{9}) to be  
\begin{align}
\omega_{\alpha}&= {1\over2}e_{\alpha a} \epsilon_{bc}
\big( e_{a}{}^{\beta} e_{b}{}^{\gamma} \partial_{\beta} e_{\gamma c}
+e_{c}{}^{\beta} e_{a}{}^{\gamma} \partial_{\beta} e_{\gamma b}
-e_{b}{}^{\beta} e_{c}{}^{\gamma} \partial_{\beta} e_{\gamma a} \big)
\nonumber \\ 
&= -\delta_{\alpha 2} \cos\theta  \,, 
\label{10}
\end{align}
where $\epsilon_{bc}$ $(\epsilon_{12}=1)$ is the Levi-Civita 
tensor in two-dimensional flat space.

In addition to these quantities, 
we now choose the gauge potential $A_{\alpha}$ to be that describing  
a monopole configuration \cite{Fel, Shn}, 
\begin{align}
A_{\alpha}=\delta_{\alpha 2} g(\kappa-\cos \theta) 
=e_{\alpha 2} \frac{g(\kappa-\cos \theta)}{r\sin \theta} \,,
\label{11}
\end{align}
where $\kappa=-1,0,1$, and $g$ denotes a magnetic charge. 
The field strength $F_{\alpha \beta}$ of this potential is 
actually the static magnetic field due to a point magnetic monopole of strength $g$ 
situated at the point $r=0\,$: 
\begin{align}
F_{\alpha\beta}=\varepsilon_{\alpha\beta} g\sin\theta 
=\epsilon_{ab} e_{\alpha a} e_{\beta b} \frac{g}{r^2} \,, 
\label{12}
\end{align} 
where $\varepsilon_{\alpha\beta}$ $(\varepsilon_{12}=1)$ is the covariant  
Levi-Civita tensor {\em density} in two dimensions.  From  
the second expression in Eq. (\ref{11}),  it is seen that 
when $\kappa=1$, $A_{a} \equiv e_{a}{}^{\alpha} A_{\alpha}$ has a singularity only at 
the south pole, $(\pi, \phi)$, on $S^2$, while when $\kappa=-1$, 
$A_{a}$ has a singularity only at the north pole, $(0, \phi)$, on $S^2$.  
This situation holds for all spheres of all possible radii $r$ 
$(0\leq r <\infty)$, and hence it follows that, 
as $r$ runs from $0$ to $\infty$, the singularity at each pole comes to form a set of  
singularities that constitutes a semi-infinite string whose endpoint is at $r=0$. 
Such a semi-infinite string is known as a Dirac string, 
and the cases $\kappa=1$ and $\kappa=-1$ are sometimes referred to as the 
{\it Dirac formalism} \cite{Fel}. By contrast, when $\kappa=0$, 
$A_{a}$ has singularities at both the north and south poles, 
in accordance with space-reflection considerations. 
As a result,  as $r$ runs from $0$ to $\infty$, these two singularities 
come to form a set of singularities that constitutes an infinite string with no ends. 
The case $\kappa=0$ is sometimes referred to as the {\it Schwinger  
formalism} \cite{Fel}, because Schwinger studied the case $\kappa=0$ in particular 
to maintain compatibility of the magnetic-charge concept with 
the principles of relativistic quantum field theory \cite{Sch}.  
Essentially, the Dirac formalism treats a semi-infinite string  
and yields the charge quantization condition $eg=n/2$, 
whereas the Schwinger formalism treats an infinite string 
and yields the charge quantization condition $eg=n$. 
In the present paper, we consider both the Dirac formalism and 
the Schwinger formalism to derive the corresponding charge-quantization 
conditions by utilizing the Atiyah-Singer index theorem in two dimensions.

Substituting Eqs. (\ref{9})--(\ref{11}) into Eq. (\ref{2}), 
we can write the massless Dirac equation $\iDsla \varphi=0$ in 
the following form \cite{Abr}: 
\begin{align}
\frac{i}{r} 
\left(\begin{array}{cc}
            0 & \nabla_{\theta}-\dfrac{i}{\sin \theta} \nabla_{\phi} \\
            \nabla_{\theta}+\dfrac{i}{\sin \theta} \nabla_{\phi} & 0 \\
          \end{array} \right) \!
\left(\begin{array}{c} 
      u^{+}(\theta, \phi) \\
      u^{-}(\theta, \phi) \\
\end{array} \right)
=\left(\begin{array}{c} 
      0 \\
      0 \\
\end{array} \right) \,,
\label{13}
\end{align}    
with 
\begin{align}
\nabla_{\theta}\equiv \frac{\partial}{\partial\theta} 
+{1\over2}\cot\theta \,, \quad 
\nabla_{\phi}\equiv \frac{\partial}{\partial\phi}
-ieg (\kappa-\cos \theta) \,. 
\label{14}
\end{align}
Equation (\ref{13}) can be separated into two differential equations 
by substituting 
$u^{\pm}(\theta, \phi)=v^{\pm}(\theta)w^{\pm}(\phi)$ into it. 
The resulting differential equation in $\phi$ can be immediately solved, 
and we obtain the normalized solutions 
$w^{\pm}_{m_{\pm}}(\phi) =(2\pi)^{-1/2}e^{im_{\pm} \phi}$. 
Here $m_{+}$ and $m_{-}$ are half-integers, that is, 
$m_{+}, m_{-}=\pm1/2,\, \pm3/2,\, \ldots$, because the spinor field 
$\varphi$ has to change sign under a $2\pi$ rotation in $\phi$.  
The differential equation in $\theta$ is thus obtained as 
\begin{align}
\left[\, \frac{d}{d\theta} + \bigg( {1\over2}\mp eg \bigg) \cot\theta 
\mp \frac{m_{\pm}-eg\kappa}{\sin\theta}\, \right]\! v^{\pm}(\theta)=0 \,.
\label{15}
\end{align}
The normalized solutions of this equation are readily found to be 
\begin{align}
v^{\pm}_{m_{\pm}}(\theta)=c_{m_{\pm}}^{\pm}
\biggl(\sin\frac{\theta}{2}\biggr)^{\!p^{\pm}_{m_{\pm}}} 
\biggl(\cos\frac{\theta}{2}\biggr)^{q^{\pm}_{m_{\pm}}} ,
\label{16}
\end{align}
where 
\begin{align}
p^{\pm}_{m_{\pm}} &\equiv \pm\{ m_{\pm}-eg(\kappa-1)\}-{1\over2}\,,  
\label{17} 
\\ 
q^{\pm}_{m_{\pm}} &\equiv \mp\{ m_{\pm}-eg(\kappa+1)\}-{1\over2}\,, 
\label{18}
\end{align}
and $c_{m_{\pm}}^{\pm}$ is the normalization constant applicable to  
$p^{\pm}_{m_{\pm}}, q^{\pm}_{m_{\pm}}>-1\,$:
\begin{align}
c_{m_{\pm}}^{\pm}
=\sqrt{ 
\frac{\varGamma(p^{\pm}_{m_{\pm}}+q^{\pm}_{m_{\pm}}+2)}
{2\varGamma(p^{\pm}_{m_{\pm}}+1) \varGamma(q^{\pm}_{m_{\pm}}+1)}
} \,.
\label{19}
\end{align}
The solution $v^{\pm}_{m_{\pm}}$ diverges at neither $\theta=0$ nor $\pi$ 
if and only if $p^{\pm}_{m_{\pm}}, q^{\pm}_{m_{\pm}}\geq 0$.  
In this case,  $v^{\pm}_{m_{\pm}}$ has a definite 
normalization constant $c_{m_{\pm}}^{\pm}$. 
The conditions $p^{+}_{m_{+}}, q^{+}_{m_{+}}\geq 0$  
necessary for $v^{+}_{m_{+}}$ to be finite at $\theta=0, \pi$ can be      
expressed as  
\begin{align}
{1\over2}+eg(\kappa-1) \leq m_{+} \leq -{1\over2}+eg(\kappa+1) \,,
\label{20}
\end{align}
implying $eg \geq 1/2$, while 
the conditions $p^{-}_{m_{-}}, q^{-}_{m_{-}}\geq 0$  
necessary for $v^{-}_{m_{-}}$ to be finite at $\theta=0, \pi$ can be      
expressed as  
\begin{align}
{1\over2}+eg(\kappa+1) \leq m_{-} \leq -{1\over2}+eg(\kappa-1) \,,
\label{21}
\end{align}
implying $eg \leq -1/2$.  
Therefore the conditions $p^{+}_{m_{+}}, q^{+}_{m_{+}}\geq 0$ 
and the conditions $p^{-}_{m_{-}}, q^{-}_{m_{-}}\geq 0$ are never satisfied  
simultaneously with a given $eg$.  
For this reason, the possible {\em regular} solutions of 
$\iDsla \varphi=0$ are restricted to 
\begin{align}
\varphi^{+}_{m_{+}}=\left(\begin{array}{c} 
      u^{+}_{m_{+}} \\
      0 \\
\end{array} \right) \;\,\mbox{for}\;\; eg \geq 1/2 \,, \quad 
\varphi^{-}_{m_{-}}=\left(\begin{array}{c} 
      0 \\ 
      u^{-}_{m_{-}} \\
      \end{array} \right) \;\,\mbox{for}\;\; eg \leq -1/2 \,,
\label{22}
\end{align}    
with $u^{\pm}_{m_{\pm}} \equiv v^{\pm}_{m_{\pm}} w^{\pm}_{m_{\pm}}$. 
They satisfy the orthonormality relation 
\begin{align}
\int_0^{\pi} \sin\theta d\theta \int_{0}^{2\pi} d\phi\, 
\big(\varphi^{\pm}_{m_{\pm}}\big)^{\dagger} 
\varphi^{S}_{m_{S}^{\prime}} 
=\left\{ \begin{array}{ll}
\delta_{{m_{\pm}}{m_{\pm}^{\prime}}}\,, & \,(S=\pm) \\ 
0 \,.& \,(S=\mp) 
\end{array}
\right. 
\label{23}
\end{align}
Each of these solutions, characterized by $m_{\pm}$, has a definite chirality, 
satisfying Eq. (\ref{6}) as well as Eq. (\ref{5}). 
Since $\sigma_3 \varphi^{\pm}_{m_{\pm}}=\pm \varphi^{\pm}_{m_{\pm}}$, 
the solutions $\{\varphi^{+}_{m_{+}}\}$ are 
recognized as  
the positive chirality zero-modes, while  
the solutions $\{\varphi^{-}_{m_{-}}\}$ are 
recognized as 
the negative chirality zero-modes. 
It is now obvious that $\frak{n}_{-}=0$ for $eg \geq 1/2$, 
$\frak{n}_{+}=0$ for $eg \leq -1/2$, and 
$\frak{n}_{+}=\frak{n}_{-}=0$ for $|eg| < 1/2$. 
When $e=0$, the Dirac operator $\iDsla$, of course, has no zero-modes.  
This illustrates the Lichnerowicz vanishing theorem \cite{Lic}. 
(Also see under Eq. (\ref{A18-2}) in the Appendix.)

\section{Count of zero-modes and 
derivation of  charge quantization conditions}   

We now count the number of zero-modes of $\iDsla$ 
to find $\frak{n}_{+}$ for $eg \geq 1/2$ and 
$\frak{n}_{-}$ for $eg \leq -1/2$, 
and then derive the charge quantization conditions.

First, consider the case $\kappa=1$,  one case of the Dirac formalism.  
When $eg\;(\geq 1/2)$ is in the interval  
$n/2 \leq eg < (n+1)/2$ with $n=1,2,3,\ldots$, 
the allowed values of $m_{+}$ can be seen immediately from 
Eq. (\ref{20}) to be $m_{+}=1/2, 3/2, \ldots, (2n-1)/2$. 
(Recall here that $m_{+}$ and $m_{-}$ take half-integer values.)  
This implies that the number of regular solutions of $\iDsla \varphi=0$ 
is $n$, and it follows that $\frak{n}_{+}=n$. 
When $eg\;(\leq -1/2)$ is in the interval $-(n+1)/2 < eg \leq -n/2$ 
with $n=1,2,3,\ldots$, the allowed values of $m_{-}$ are found from 
Eq. (\ref{21}) to be $m_{-}=-1/2, -3/2, \ldots, -(2n-1)/2$, and   
it follows that $\frak{n}_{-}=n$. 
The total magnetic flux due to a point magnetic monopole 
is obtained from Eq. (\ref{12}) as  
\begin{align}
{1\over2} \int_{S^2} d\theta d\phi \, 
\varepsilon^{\alpha\beta} F_{\alpha\beta}
=g \int_0^{\pi} \sin\theta d\theta \int_0^{2\pi} d\phi =4\pi g \,.
\label{24}
\end{align}
Thus, in this case, the Atiyah-Singer index theorem (\ref{7-1}) leads to 
the following relations: 
$n=2eg$ for $eg\geq 1/2$,  $-n=2eg$ for $eg\leq -1/2$, and 
$0=2eg$ for $|eg|<1/2$. 
These are compatible with the choice of the intervals for $eg$ mentioned above.  
More precisely, because the relation $n=2eg$ ($-n=2eg$) 
was derived for $eg$ assumed 
to be in the interval $n/2 \leq eg < (n+1)/2\,$ ($-(n+1)/2 < eg \leq -n/2$) 
closed at $eg=n/2$ ($eg=-n/2$),    
the compatibility of the relation and the interval is maintained. 
The three relations found here are brought together in the form $eg=n/2$ with 
$n\in \Bbb Z$.  This is precisely the Dirac quantization condition.

Next, consider the case $\kappa=-1$, which is another case of the Dirac formalism, 
though it is a mirror image of the case $\kappa=1$. 
When $eg\;(\geq 1/2)$ is in the interval  
$n/2 \leq eg < (n+1)/2$ with $n=1,2,3,\ldots$, 
the allowed values of $m_{+}$ are found from Eq. (\ref{20}) to be 
$m_{+}=-1/2, -3/2, \ldots, -(2n-1)/2$, so that $\frak{n}_{+}=n$. 
When $eg\;(\leq -1/2)$ is in the interval $-(n+1)/2 < eg \leq -n/2$ 
with $n=1,2,3,\ldots$, the allowed values of $m_{-}$ are found from 
Eq. (\ref{21}) to be $m_{-}=1/2, 3/2, \ldots, (2n-1)/2$, so that 
$\frak{n}_{-}=n$. Therefore the possible values of $\frak{n}_{+}$ and  
$\frak{n}_{-}$ are the same as those in the case $\kappa=1$, as might  
be expected; the Dirac quantization condition follows again, maintaining  
the compatibility with the choice of the intervals for $eg$.

Finally, consider the case $\kappa=0$, i.e., that of the Schwinger formalism. 
When $eg$ is in the interval $1/2 \leq eg <1$, there are no allowed values of 
$m_{+}$, as can be seen from Eq. (\ref{20}), so that $\frak{n}_{+}=0$.  
When $eg\;(\geq 1)$ is in the interval  
$n \leq eg < n+1$ with $n=1,2,3,\ldots$, 
the allowed values of $m_{+}$ can be seen immediately from 
Eq. (\ref{20}) to be $m_{+}=\pm1/2, \pm3/2, \ldots, \pm(2n-1)/2$, and 
it follows that $\frak{n}_{+}=2n$. 
When $eg$ is in the interval $-1< eg \leq -1/2$, there are no allowed values of 
$m_{-}$, as can be seen from Eq. (\ref{21}), so that $\frak{n}_{-}=0$.  
When $eg\;(\leq -1)$ is in the interval  
$-(n+1)< eg \leq -n$ with $n=1,2,3,\ldots$, 
the allowed values of $m_{-}$ are found from 
Eq. (\ref{21}) to be $m_{-}=\pm1/2, \pm3/2, \ldots, \pm(2n-1)/2$, and 
it follows that $\frak{n}_{-}=2n$. 
Thus, in the case $\kappa=0$, the Atiyah-Singer index theorem (\ref{7-1})  
leads to the following relations: 
$2n=2eg$ for $eg\geq 1$,  $-2n=2eg$ for $eg\leq -1$, and 
$0=2eg$ for $|eg|<1$. 
These are also compatible with the choice of the intervals for $eg$. 
The three relations are brought together in the form $eg=n$  
with $n\in \Bbb Z$. 
This is precisely the Schwinger quantization condition.

\section{Conclusions}

We have proposed a novel method for deriving the charge quantization conditions 
that are caused by the presence of a point magnetic monopole situated at the origin in 
three-dimensional Euclidean space. 
Although our investigation considers four-dimensional spacetime, 
the problem itself is reduced to one in two-dimensional space by virtue of 
the static and scale-independent conditions of the system 
[see Eqs. (\ref{11}) and (\ref{12})]. 
For this reason, it was possible to examine the charge quantization conditions 
by placing two-component static chiral spinor fields on a sphere at whose centre 
there exists a point magnetic monopole.

Employing such a system,  
we derived both the Dirac quantization condition, $eg=n/2$ ($n \in \Bbb Z$),  
and the Schwinger quantization condition, $eg=n$ ($n \in \Bbb Z$),  
by utilizing the Atiyah-Singer index theorem in two dimensions. 
These quantization conditions can be understood as   
the conditions that the Atiyah-Singer index theorem in two dimensions 
holds without giving rise to the index defect.  
In other words, the charge quantization conditions may be regarded as the 
consistency conditions to be satisfied when zero-modes of 
the Dirac operator are allowed to exist on a sphere at whose centre there lies    
a point magnetic monopole.

As we have seen, the Dirac quantization condition, $eg=n/2$, is found when $\kappa$ 
in Eq. (\ref{11}) is equal to $1$ or $-1$, 
while the Schwinger quantization condition, $eg=n$, is found when $\kappa=0$. 
These two conditions were obtained by solving the simple Dirac equation (\ref{13})   
and by formally counting the number of zero-modes of the Dirac operator.   
In our procedure, the difference between the Dirac and Schwinger quantization 
conditions simply results from the fact that 
the number of zero-modes in the Schwinger formalism is twice that  
in the Dirac formalism for a fixed integer $n$.

Our approach requires neither a careful treatment of a string singularity 
in the gauge potential nor the concept of patches (or charts) and sections.  
In fact, in this paper, we have treated neither the gauge transformation of  
wave functions (or sections) nor of gauge potentials.

Finally, note that if we take the charge quantization conditions  
$eg=n/2$ and $eg=n$ as given by other approaches, 
the argument presented in this paper may be understood as an illustration or verification 
of the Atiyah-Singer index theorem in a particular case.

\appendix
\section{}

In this appendix, 
we prove the Atiyah-Singer index theorem in two dimensions,   
expressed by Eq. (\ref{7}), by using the heat kernel method \cite{Avr,Ber,Par,BB,Vas}.

Let us consider the eigenvalue equation for the Dirac operator $\iDsla$ 
given in Eq. (\ref{2}), 
\begin{align}
\iDsla \varphi_N (q)=\lambda_n  \varphi_N (q) \,, 
\label{A1}
\end{align}
with an eigenvalue $\lambda_n$ and an eigenfunction $\varphi_N$. 
Here, $N$ is a collective index representing the pair of indices $(n,\nu)$,  
where $n$ labels the eigenvalues of $\iDsla$ in such a way that  
$n\neq n'\,\, \rightleftarrows \,\, \lambda_n \neq \lambda_{n'}$, 
while $\nu$ distinguishes between the degenerate eigenfunctions of $\iDsla$ 
corresponding to the same eigenvalue $\lambda_n$. 
Because $\iDsla$ is self-adjoint, as can be proven by using the premise 
that $\mathcal M$ has no boundary, the eigenvalue $\lambda_n$ is real and 
the eigenfunctions $\{\varphi_N \}$ form an orthonormal set: 
\begin{align}
\int_\mathcal{M} d^2 q \sqrt{\frak{g}(q)}\,  
\varphi^{\dagger}_N (q) \varphi_{N'} (q) 
=\delta_{NN'} \, , 
\label{A2}
\end{align}
with $\sqrt{\frak{g}} \equiv |\det(e_{a}{}^{\alpha})|^{-1}$ and 
$\delta_{NN'}\equiv \delta_{nn'}\delta_{\nu\nu'}$. 
(Because $\mathcal{M}$ is compact, 
the normalization of $\varphi_N $ is always possible.) 
The following completeness condition can be imposed on $\{\varphi_N \}$: 
\begin{align}
\sum_{N}  \varphi_N (q) \varphi^{\dagger}_N (q')
=\delta^{2}(q, q') \sigma_0 
\,. 
\label{A3}
\end{align}
Here, $\delta^{2}(q,q')$ is a generalized delta-function on 
two-dimensional curved space \cite{Avr,FS}, 
satisfying 
\begin{align}
f(q)=\int_\mathcal{M} d^2 q' 
\sqrt{\frak{g}(q')}\, \delta^{2}(q, q') f(q') 
\label{A4}
\end{align}
for an arbitrary smooth function $f(q)$ on $\mathcal{M}$.\footnote
{On the right-hand side of Eq. (\ref{A3}), 
the appearance of the unit matrix $T_0$ in 
a representation space of the Lie group $\mathcal{G}$, 
satisfying $(T_0\otimes \sigma_0) \varphi_N (q)=\varphi_N (q)$, is understood.
The appearance of this matrix is also understood in some equations in this paper.}

Let us now consider the function  
\begin{align}
\mathcal{A}(q) \equiv \sum_{N} \varphi^{\dagger}_N (q) \sigma_3 \varphi_N (q) \,.
\label{A5} 
\end{align} 
This function is not well-defined, because it involves an infinite sum of the functions 
$\varphi^{\dagger}_N \sigma_3 \varphi_N$ at the same point $(q^\alpha)$. 
Hence we have to carry out a suitable 
regularization of $\mathcal{A}(q)$ in order to evaluate it.  
For this purpose, we insert 
a Gaussian cutoff with a positive parameter $\tau$ into the right-hand side of Eq. (\ref{A5}) 
and rewrite the regularized form  
by using the eigenvalue equation (\ref{A1}):  \cite{Ber, FS}
\begin{align}
\mathcal{A}_{\rm reg} (q) & \equiv 
\lim_{\tau\searrow 0} 
\sum_{N} \varphi^{\dagger}_N (q) \sigma_3 e^{-\tau \lambda_N^2} \varphi_N (q) 
\nonumber \\ 
& =\lim_{\tau\searrow 0} 
\sum_{N} \varphi^{\dagger}_N (q) \sigma_3 
\exp\!\big[-\tau (\iDsla)^2 \big]  \varphi_N (q) 
\nonumber \\ 
& =\lim_{\tau\searrow 0} 
 {\rm Tr}\Big( \sigma_3 \sum_{N} \big\{ \exp\!\big[-\tau (\iDsla)^2 \big] 
\varphi_N (q)\big\} \varphi^{\dagger}_N (q) \Big)
\nonumber \\
& =\lim_{\tau\searrow 0} \lim_{q'\rightarrow q}
 {\rm Tr}\big( \sigma_3 G(q,q',\tau) \big)
\,,
\label{A6} 
\end{align} 
where the notation ^^ ^^ Tr" indicates the trace taken over the Pauli matrices and over 
the generators of the Lie group $\mathcal{G}$. 
The two-point function $G$ is defined by 
\begin{align}
G(q,q',\tau)\equiv \sum_{N} \big\{ \exp\!\big[-\tau (\iDsla)^2 \big] 
\varphi_N (q)\big\} \varphi^{\dagger}_N (q') 
\,.
\label{A7}
\end{align} 
It is readily shown that $G$ satisfies the so-called heat equation,  
\begin{align}
-\frac{\partial}{\partial \tau}G(q,q',\tau)
=(\iDsla)^2 G(q,q',\tau) \,,
\label{A8}
\end{align}
as well as the initial condition due to the completeness condition (\ref{A3}), 
\begin{align}
G(q,q',0)=\delta^{2}(q,q') \sigma_0 
\,.
\label{A9} 
\end{align}
The function $G(q,q',\tau)$ is often referred to as the heat kernel.

Before trying to solve Eq. (\ref{A8}) supplemented with Eq. (\ref{A9}), 
we calculate the square of the Dirac operator $\iDsla$. 
To this end, it is convenient to introduce the affine connection  
$\varGamma_{\alpha\beta}{}^{\gamma}$, which appears, for instance, 
in the condition describing the parallel transport on $\mathcal{M}$,  
\begin{align}
\partial_{\beta} e_{a}{}^{\alpha}+\varGamma_{\beta\gamma}{}^{\alpha} e_{a}{}^{\gamma}
+\epsilon_{ab}\omega_{\beta} e_{b}{}^{\alpha}=0 \,,
\label{A10}
\end{align}
where $\epsilon_{ab}$ $(\epsilon_{12}=1)$ is the Levi-Civita tensor in two-dimensional 
flat space. Because $\mathcal{M}$ is a Riemannian manifold and thus is torsion-free, 
the affine connection is symmetric in the lower indices: 
$\varGamma_{\beta\gamma}{}^{\alpha}=\varGamma_{\gamma\beta}{}^{\alpha}$. From 
this relation and Eq. (\ref{A10}), 
the affine connection is determined in terms of the inverse metric 
$g^{\alpha\beta}= \delta_{ab}e_{a}{}^{\alpha} e_{b}{}^{\beta}$ 
and the derivative of the metric 
$g_{\alpha\beta}= \delta_{ab} e_{\alpha a} e_{\beta b}$  
to be 
\begin{align}
\varGamma_{\alpha\beta}{}^{\gamma}={1\over2} g^{\gamma\delta}
(\partial_{\alpha} g_{\beta\delta} +\partial_{\beta} g_{\delta\alpha}
-\partial_{\delta} g_{\alpha\beta}) \,.
\label{A11}
\end{align}
Also, the quadratic operator $(\iDsla)^2$ can be calculated, using the fundamental 
properties of the Pauli matrices     
$[\sigma_i, \sigma_j]=2i\epsilon_{ijk}\sigma_k$ and  
$\{\sigma_i, \sigma_j\}=2\delta_{ij}\,$,  
as 
\begin{align}
(\iDsla)^2
=-\sigma_0 g^{\alpha\beta} 
(D_\alpha D_\beta -\varGamma_{\alpha\beta}{}^{\gamma}D_\gamma) 
-\frac{1}{4} [\sigma^\alpha, \sigma^\beta]  [D_\alpha, D_\beta] \,,
\label{A12}
\end{align}
where $\sigma^\alpha \equiv \sigma_a e_{a}{}^{\alpha}$. 
The commutator of the covariant derivatives is easily calculated  
from Eq. (\ref{3}) as 
\begin{align}
[D_\alpha, D_\beta]=\frac{i}{2} \varOmega_{\alpha\beta}\sigma_3 
-ieF_{\alpha\beta}\sigma_0 \,,
\label{A13}
\end{align}
with
\begin{align}
\varOmega_{\alpha\beta}&\equiv 
\partial_\alpha \omega_{\beta}-\partial_\beta \omega_{\alpha} \,, 
\label{A14}
\\
F_{\alpha\beta}&\equiv \partial_{\alpha}A_{\beta}-\partial_{\beta}A_{\alpha}
-ie[A_{\alpha}, A_{\beta}] \,.
\label{A15}
\end{align}
Substituting Eq. (\ref{A13}) into Eq. (\ref{A12}), we can proceed with calculating 
the quadratic operator, and we finally obtain   
\begin{align}
(\iDsla)^2=-\sigma_0 \Delta +\frac{1}{4}R \sigma_0 
-\frac{1}{2}e \epsilon^{\alpha\beta} F_{\alpha\beta} \sigma_3 \,,
\label{A16}
\end{align}
with 
\begin{align}
\Delta & \equiv   
g^{\alpha\beta}(D_\alpha D_\beta -\varGamma_{\alpha\beta}{}^{\gamma}D_\gamma)
=\frac{1}{\sqrt{\frak{g}}} 
D_\alpha \sqrt{\frak{g}} \,g^{\alpha\beta} D_\beta \,,
\label{A17}
\\
R & \equiv \epsilon^{\alpha\beta} \varOmega_{\alpha\beta} \,, 
\label{A18}
\\
\epsilon^{\alpha\beta} & \equiv \epsilon_{ab}e_{a}{}^{\alpha} e_{b}{}^{\beta}
=\frac{1}{\sqrt{\frak{g}}} \varepsilon^{\alpha\beta}
 \,.
\label{A18-2}
\end{align}
The operator $\Delta$, which is negative-semidefinite, 
is the Laplacian taken into account the covariance under general coordinate 
transformations as well as that under local Lorentz and the extra gauge 
transformations. 
The function $R$ is the scalar curvature of $\mathcal{M}$.  
Equation (\ref{A16}) implies that for manifolds with positive curvature, 
there exist no zero modes of the Dirac operator $\iDsla$ if $e=0$. 
This is known as the Lichnerowicz vanishing theorem \cite{Lic}.

Now we return to the heat equation (\ref{A8}). 
Because the double limit $q'\rightarrow q$, $\tau\searrow 0$ 
is taken in Eq. (\ref{A6}), only the asymptotic form of $G(q,q',\tau)$ 
in this limit is necessary for the evaluation of $\mathcal{A}_{\rm reg}(q)$. 
In order to simply derive this asymptotic form,  
we employ the Riemann normal coordinates 
with the origin at $(q^{\prime}{}^\alpha)$, characterized by 
$\partial_{\gamma}g_{\alpha\beta}(q^{\prime})=0$ 
[or equivalently $\varGamma_{\alpha\beta}{}^{\gamma}(q^{\prime})=0\,$]  
and $g_{\alpha\beta}(q^{\prime})=\delta_{\alpha\beta}$ \cite{FS, Par, Hat}. 
Then, in an infinitesimal neighborhood of $(q^{\prime}{}^\alpha)$, 
denoted by $\mathcal{U}_{q^{\prime}}$, 
the metric $g_{\alpha\beta}$ at $(q^\alpha) \in \mathcal{U}_{q^{\prime}}$ 
and its inverse $g^{\alpha\beta}$ can be expressed in the forms of their respective  
Taylor series:  
\begin{align}
g_{\alpha\beta}(q)&
=\delta_{\alpha\beta}-{1\over3} R_{\alpha\gamma\beta\delta}(q^{\prime})
(q^\gamma -q^{\prime}{}^{\gamma}) (q^\delta -q^{\prime}{}^{\delta})
+O\big( (q-q')^{3} \big) \,,
\label{A19}
\\
g^{\alpha\beta}(q)&
=\delta^{\alpha\beta}
+{1\over3} R^{\alpha}{}_{\gamma}{}^{\beta}{}_{\delta}(q^{\prime})
(q^\gamma -q^{\prime}{}^{\gamma}) (q^\delta -q^{\prime}{}^{\delta})
+O\big( (q-q')^{3} \big) 
\,.
\label{A20}
\end{align}
Here, $R_{\alpha\gamma\beta}{}^{\delta}$ denotes the Riemann tensor,  
\begin{align}
R_{\alpha\gamma\beta}{}^{\delta}&=-\partial_{\alpha} \varGamma_{\gamma\beta}{}^{\delta}
+\partial_{\gamma} \varGamma_{\alpha\beta}{}^{\delta}
-\varGamma_{\alpha\epsilon}{}^{\delta} \varGamma_{\gamma\beta}{}^{\epsilon}
+\varGamma_{\gamma\epsilon}{}^{\delta} \varGamma_{\alpha\beta}{}^{\epsilon}
\nonumber \\
&= \varOmega_{\alpha\gamma} \epsilon_{ab} e_{\beta a} e_{b}{}^{\delta} 
\,, 
\label{A21}
\end{align}
with the (anti)symmetric properties 
$R_{\alpha\gamma\beta\delta}=-R_{\gamma\alpha\beta\delta}
=R_{\beta\delta\alpha\gamma}$. 
All the expansion coefficients in the two Taylor series  
(\ref{A19}) and (\ref{A20}) are written in terms of the Riemann tensor and 
its covariant derivatives of first and higher orders that are all evaluated at 
$(q^{\prime}{}^{\alpha})$ \cite{Hat}. The Taylor expansion of 
$\varGamma_{\alpha\beta\gamma}(q)$ 
about $(q^{\prime}{}^{\alpha})$ is found from Eqs. (\ref{A11}) and 
(\ref{A19}) to be 
\begin{align}
\varGamma_{\alpha\beta\gamma}(q)
=\frac{1}{3} \big( R_{\alpha\gamma\beta\delta}(q')
+R_{\beta\gamma\alpha\delta}(q') \big)
(q^\delta -q^{\prime}{}^{\delta})
+O\big( (q-q')^{2} \big) \,.
\label{A22}
\end{align}
After substituting Eqs. (\ref{A20}) and (\ref{A22}) into Eq. (\ref{A17}), 
the covariant Laplacian $\Delta$ can be written in the following form: 
\begin{align}
\Delta& = 
\bigg(\delta^{\alpha\beta}
+{1\over3} R^{\alpha}{}_{\gamma}{}^{\beta}{}_{\delta}(q^{\prime})
(q^\gamma -q^{\prime}{}^{\gamma}) (q^\delta -q^{\prime}{}^{\delta})
\bigg) D_{\alpha} D_{\beta}
\nonumber 
\\
& \quad\;\;
-\frac{2}{3} R_{\alpha}{}^{\gamma\alpha}{}_{\delta}(q^{\prime})
(q^\delta -q^{\prime}{}^{\delta}) D_{\gamma}
+O\big( (q-q')^{3} \big) \,.
\label{A23}
\end{align}

For the special case  
$(\iDsla)^2 =-\sigma_0 \delta^{\alpha\beta}
\partial_{\alpha} \partial_{\beta}$, supplemented with 
the condition 
\begin{align}
G(q,q',0)=\delta^{2}(q,q')\sigma_0
=\delta(q^{1}-q^{\prime}{}^{1})\delta(q^{2}-q^{\prime}{}^{2})\sigma_0 \,, 
\quad (q^{\alpha})\in \mathcal{U}_{q'} \,,
\label{A24}
\end{align}
the heat equation (\ref{A8}) has the Gaussian solution 
\begin{align}
\widetilde{G} (q,q',\tau)=
\frac{1}{4\pi \tau}\sigma_0  \exp\! \left[ -\frac{1}{4\tau}
\delta_{\alpha\beta}
(q^{\alpha}-q^{\prime}{}^{\alpha}) (q^{\beta}-q^{\prime}{}^{\beta}) \right] .
\label{A25}
\end{align}
We are now interested in the behavior of $G(q,q',\tau)$ 
for extremely small $\tau$. In that case, the following expansion is valid: 
\begin{align}
G (q,q',\tau)=\widetilde{G} (q,q',\tau) 
\sum_{k=0}^{\infty} a_{k}(q,q')\tau^{k} \,, 
\quad 0< \tau \ll 1 \,, 
\label{A26}
\end{align} 
where $a_k (q,q')$ $(k=0,1,2,\ldots)$ are the so-called De Witt--Seeley 
coefficients \cite{Avr}. Substituting Eq. (\ref{A26}) into the heat equation (\ref{A8})  
with the covariant Laplacian (\ref{A23}), we obtain 
the recursion relations  
\begin{align}
&\left\{ k+(q^{\alpha}-q^{\prime}{}^{\alpha}) D_{\alpha}
-\frac{1}{6} R^{\gamma}{}_{\alpha\gamma\beta}(q^{\prime})
(q^{\alpha}-q^{\prime}{}^{\alpha}) (q^{\beta}-q^{\prime}{}^{\beta}) 
+O\big( (q-q')^{3} \big) \right\} a_k (q, q')
\nonumber \\ 
&=
-\left \{ (\iDsla)^2 +O\big( (q-q')^{3} \big) \right\} a_{k-1}(q, q')  \,, 
\qquad k=1,2,3, \ldots 
\label{A27}
\end{align}
and
\begin{align}
\left \{ (q^{\alpha}-q^{\prime}{}^{\alpha}) D_{\alpha}
-\frac{1}{6} R^{\gamma}{}_{\alpha\gamma\beta}(q^{\prime})
(q^{\alpha}-q^{\prime}{}^{\alpha}) (q^{\beta}-q^{\prime}{}^{\beta}) 
+O\big( (q-q')^{3} \big) \right\} a_0 (q, q')=0 \,. 
\label{A28}
\end{align}
In the derivation of these relations, some irrelevant terms vanish  
by virtue of the antisymmetric property of the indices of the Riemann tensor. 
Then, taking the covariant derivative of Eq. (\ref{A28}) yields  
\begin{align}
\left \{ D_{\beta} +(q^{\gamma}-q^{\prime}{}^{\gamma}) 
\bigg( D_{\beta} D_{\gamma} -\frac{1}{3} 
R^{\delta}{}_{\beta\delta\gamma}(q^{\prime}) \bigg) 
+O\big( (q-q')^{2} \big) \right\} a_0 (q, q')=0 \,. 
\label{A29}
\end{align}
Next, applying $D_{\alpha}$ to this formula   
and contracting the indices $\alpha$ and $\beta$, we have   
\begin{align}
\left\{ g^{\alpha\beta} D_{\alpha} D_{\beta} 
-\frac{1}{6} R(q^{\prime}) +O(q-q') \right\} a_0 (q, q')=0 \,, 
\label{A30}
\end{align}
where $R=g^{\alpha\beta}R^{\gamma}{}_{\alpha\gamma\beta}$ is the scalar curvature 
which, of course, is identical to $R$ defined in Eq. (\ref{A18}), 
as can be seen from Eq. (\ref{A21}).

It is easy to see that the condition (\ref{A24}) yields the normalization condition 
\begin{align}
\lim_{q' \rightarrow q} a_0 (q, q')=\sigma_0 \,. 
\label{A31}
\end{align}
In addition, it follows from Eqs. (\ref{A29}) and (\ref{A30}) that 
\begin{align}
&\lim_{q' \rightarrow q} D_{\beta} a_0 (q, q')=0 \,, 
\label{A32}
\\
&\lim_{q' \rightarrow q} g^{\alpha\beta} D_{\alpha} D_{\beta} a_0 (q, q') 
=\frac{1}{6} R \sigma_0 \,. 
\label{A33}
\end{align}
Taking the limit $q'\rightarrow q$ in the recursion relation (\ref{A27}) 
with $k=1$ and recalling Eqs. (\ref{A16}) and (\ref{A17}), we have  
\begin{align}
&\lim_{q' \rightarrow q} a_1 (q, q')
=-\lim_{q' \rightarrow q} (\iDsla)^2 a_0 (q, q')
\nonumber \\
&= \lim_{q' \rightarrow q} \left\{ 
\sigma_0 g^{\alpha\beta}(D_\alpha D_\beta 
-\varGamma_{\alpha\beta}{}^{\gamma}D_\gamma) -\frac{1}{4}R\sigma_0 
+\frac{1}{2}e \epsilon^{\alpha\beta} F_{\alpha\beta} \sigma_3
\right\} a_{0}(q,q^{\prime})
\nonumber \\
&=-{1\over12} R\sigma_0 
+\frac{1}{2}e \epsilon^{\alpha\beta} F_{\alpha\beta} \sigma_3 \,,
\label{A34}
\end{align}
where the conditions (\ref{A31})--(\ref{A33}) have been used. 
Having obtained $a_0 (q, q')$ and $a_1 (q, q')$ in the limit $q'\rightarrow q$, 
the asymptotic form of $G (q,q',\tau)$ in this limit is immediately 
found from Eq. (\ref{A26}) to be 
\begin{align}
\lim_{q' \rightarrow q} G (q,q',\tau)
= \frac{1}{4\pi\tau}\sigma_0 
-{1\over48\pi} R\sigma_0 
+\frac{1}{8\pi} e \epsilon^{\alpha\beta} F_{\alpha\beta} \sigma_3  
&+O(\tau)  \,,
\label{A35}
\\ 
& \!\!\!\!\!\!  0< \tau \ll 1 \,. 
\nonumber 
\end{align}
Next, we insert Eq. (\ref{A35}) into the last line of Eq. (\ref{A6}). 
Then, the first and second terms on the right-hand side of Eq. (\ref{A35}) 
vanish, from Eq. (\ref{A6}), when the trace of the Pauli matrices is calculated. 
After taking the limit $\tau\searrow 0$, only the third term survives, and we have   
\begin{align}
\mathcal{A}_{\mathrm{reg}}(q)=
\frac{e}{4\pi} {\rm tr}\, \epsilon^{\alpha\beta} F_{\alpha\beta} \,, 
\label{A36}
\end{align}
where the trace over the generators of $\mathcal{G}$ remains.

Let us next consider the chiral decomposition of the eigenfunction $\varphi_N$ 
of the Dirac operator $\iDsla$ \cite{Ber, FS},  
\begin{align}
\varphi_{N}^{\pm}\equiv \frac{1}{2} (\sigma_0 \pm \sigma_3 ) \varphi_N \,, \,
\quad \sigma_3 \varphi_{N}^{\pm}=\pm \varphi_{N}^{\pm} \,.
\label{A37}
\end{align}
The positive chirality component $\varphi_{N}^{+}$ and the negative chirality 
component $\varphi_{N'}^{-}$ are orthogonal in the sense that 
$\varphi_{N}^{+\,\dagger} \varphi^{-}_{N'}
=\varphi_{N'}^{-\,\dagger} \varphi^{+}_{N}=0$.  
In terms of $\varphi_{N}^{\pm}$, due to the relation $\{\iDsla,\, \sigma_3 \}=0$, 
the eigenvalue equation (\ref{A1}) can be written as 
\begin{align}
\iDsla \varphi_N^{\pm} (q)=\lambda_n  \varphi_N^{\mp} (q) \,.
\label{A38}
\end{align}
Now we assume that $\lambda_0 =0\,$; accordingly, the corresponding eigenfunctions   
$\varphi_{0,\nu_{\pm}}^{\pm}$ are treated as the zero-modes of $\iDsla$.  
Except for $\varphi_{0,\nu_{\pm}}^{\pm}$, 
the positive and negative chirality components are not eigenfunctions 
of $\iDsla$. Equation (\ref{A38}) shows that 
when $n\neq0$, or equivalently $\lambda_n \neq 0$, 
there is a one-to-one correspondence between $\varphi_{N}^{+}$ 
and $\varphi_{N}^{-}$. Consequently, it follows that 
the number of positive chirality components 
$\{\varphi_{N}^{+}\}_{n\neq0}$ is equal to that of 
negative chirality components $\{\varphi_{N}^{-}\}_{n\neq0}$.   
Also, when $n \neq 0$, the relation  
\begin{align}
\int_\mathcal{M} d^2 q \sqrt{\frak{g}(q)}\,  
\varphi_{N}^{+\,\dagger} (q) \varphi^{+}_{N} (q) 
=\int_\mathcal{M} d^2 q \sqrt{\frak{g}(q)}\,  
\varphi_{N}^{-\,\dagger} (q) \varphi^{-}_{N} (q)  
\label{A39}
\end{align}
can be proved using Eq. (\ref{A38}), together with the self-adjointness of $\iDsla$ 
and the fact that $\lambda_n$ is purely real. 
Since the zero-modes $\varphi_{0,\nu_{\pm}}^{\pm}$ are eigenfunctions of $\iDsla$, 
the set of zero-modes $\{\varphi_{0,\nu_{\pm}}^{\pm}\}$ with 
the orthonormality condition 
\begin{align}
\int_\mathcal{M} d^2 q \sqrt{\frak{g}(q)}\, 
\varphi_{0,\nu_{\pm}}^{\pm\,\dagger}(q) 
\varphi_{0,\nu_{S}^{\prime}}^{S}(q) 
=\left\{ \begin{array}{ll}
\delta_{{\nu_{\pm}}{\nu_{\pm}^{\prime}}} \,, & \,(S=\pm) \\ 
0 \,, & \,(S=\mp) 
\end{array}
\right. 
\label{A40}
\end{align}
can be taken as a subset of the orthonormal set $\{ \varphi_N \}$. 
In Eq. (\ref{A40}), the orthogonality relation in the case $S=\mp$ is valid 
owing to the condition $\varphi_{0,\nu_{\pm}}^{\pm\,\dagger} 
\varphi_{0,\nu_{\mp}^{\prime}}^{\mp}=0$,  
which is derived from the eigenvalue equation 
$\sigma_3 \varphi_{0,\nu_{\pm}}^{\pm}=\pm \varphi_{0,\nu_{\pm}}^{\pm}$.

Using Eqs. (\ref{A39}) and (\ref{A40}), 
the integral of $\mathcal{A}_{\rm reg} (q)$ over $\mathcal{M}$ can be evaluated 
from the first line of Eq. (\ref{A6}), and we find 
\begin{align}
&\int_\mathcal{M} d^2 q \sqrt{\frak{g}(q)}\, \mathcal{A}_{\rm reg} (q) 
\nonumber \\ 
& =\lim_{\tau\searrow 0} \sum_{N} e^{-\tau \lambda_N^2}
\int_\mathcal{M} d^2 q \sqrt{\frak{g}(q)}\, 
\big( \varphi_{N}^{+}{}^{\dagger} (q) \varphi^{+}_{N} (q) 
-\varphi_{N}^{-}{}^{\dagger} (q) \varphi^{-}_{N} (q) \big) 
\nonumber \\ 
&=\sum_{\nu_{+}} \int_\mathcal{M} d^2 q \sqrt{\frak{g}(q)}\, 
\varphi_{0,\nu_{+}}^{+\,\dagger}(q) \varphi_{0,\nu_{+}}^{+}(q) 
-\sum_{\nu_{-}} \int_\mathcal{M} d^2 q \sqrt{\frak{g}(q)}\, 
\varphi_{0,\nu_{-}}^{-\,\dagger}(q) \varphi_{0,\nu_{-}}^{-}(q) 
\nonumber \\ 
&=\frak{n}_{+}-\frak{n}_{-} \,, 
\label{A41}
\end{align} 
where $\frak{n}_{+}\equiv\sum_{\nu_{+}} 1$  
$\big(\, \frak{n}_{-}\equiv\sum_{\nu_{-}} 1 \, \big)$   
represents the number of positive (negative) chirality zero-modes. 
Combining Eqs. (\ref{A36}) and (\ref{A41}) and noting Eq. (\ref{A18-2}), 
we have  
\begin{align}
\frak{n}_{+}-\frak{n}_{-}=\frac{e}{4\pi} \int_\mathcal{M} d^2 q \,
{\rm tr}\, \varepsilon^{\alpha\beta} F_{\alpha\beta} \,. 
\label{A42}
\end{align}
Thus, the Atiyah-Singer index theorem in two dimensions,   
expressed by Eq. (\ref{7}), is proven.

\end{document}